\documentclass[%
reprint,
superscriptaddress,
aps,
prl,
portrait,
]{revtex4-2} 

\usepackage{blindtext}
\usepackage{amsmath}
\usepackage{bm}
\usepackage{xcolor}
\usepackage{soul}
\usepackage{siunitx}
\usepackage[version=4]{mhchem} 
\usepackage{graphicx} 
\usepackage{subfig} 
\usepackage[normalem]{ulem}
\usepackage{pdfpages}
\graphicspath{ {./figures/} }

\usepackage{caption}

\makeatletter
\AtBeginDocument{\let\LS@rot\@undefined}
\makeatother

\begin{document}

\title{Gate Electrodes Enable Tunable Nanofluidic Particle Traps} 

\author{Philippe M. Nicollier}
\affiliation{IBM Research Europe - Zurich, S\"aumerstrasse 4, CH-8803 R\"uschlikon, Switzerland}
\author{Aaron D. Ratschow}
\affiliation{Institute for Nano- and Microfluidics, TU Darmstadt, Alarich-Weiss-Strasse 10, D-64287
Darmstadt, Germany}
\author{Francesca Ruggeri}
\affiliation{IBM Research Europe - Zurich, S\"aumerstrasse 4, CH-8803 R\"uschlikon, Switzerland}
\author{Ute Drechsler}
\affiliation{IBM Research Europe - Zurich, S\"aumerstrasse 4, CH-8803 R\"uschlikon, Switzerland}
\author{Steffen Hardt}
\affiliation{Institute for Nano- and Microfluidics, TU Darmstadt, Alarich-Weiss-Strasse 10, D-64287
Darmstadt, Germany}
\author{Federico Paratore}
\email{fparatore@ethz.ch}
\affiliation{IBM Research Europe - Zurich, S\"aumerstrasse 4, CH-8803 R\"uschlikon, Switzerland}
\affiliation{Laboratory for Soft Materials and Interfaces, Department of Materials, ETH Zürich, Vladimir-Prelog-Weg 5, CH-8093 Zürich, Switzerland}
\author{Armin W. Knoll}
\email{ark@zurich.ibm.com}
\affiliation{IBM Research Europe - Zurich, S\"aumerstrasse 4, CH-8803 R\"uschlikon, Switzerland}

\date{\today}

\begin{abstract}
The ability to control the location of nanoscale objects in liquids is essential for fundamental and applied research from nanofluidics to molecular biology. To overcome their random Brownian motion, the electrostatic fluidic trap creates local minima in potential energy by shaping electrostatic interactions with a tailored wall topography. However, this strategy is inherently static -- once fabricated the potential wells cannot be modulated. Here, we propose and experimentally demonstrate that such a trap can be controlled through a buried gate electrode.
We measure changes in the average escape times of nanoparticles from the traps to quantify the induced modulations of $0.7k_\mathrm{B}T$ in potential energy and \SI{50}{mV} in surface potential. Finally, we summarize the mechanism in a parameter-free predictive model, including surface chemistry and electrostatic fringing, that reproduces the experimental results. Our findings open a route towards real-time controllable nanoparticle traps.
\end{abstract}

\maketitle

The building blocks of life are biological objects at the nanoscale \cite{Marth_2008}. The ability to control the location of sub-micron objects in nanofluidic systems is thus of immediate importance for fundamental and applied research ranging from molecular biology to diagnostic devices. Such objects undergo thermal fluctuations, known as Brownian motion \cite{Brown1828}, that scale inversely with the particle radius. Techniques exist to manipulate small particles in liquids, but they generally rely on externally applied fields or gradients, which have an unfavorable scaling with particle volume. In an attempt to overcome this fundamental limitation, Krishnan et al. \cite{Krishnan10nature} introduced a trapping strategy called the \textit{electrostatic fluidic trap}. Their technique relies on the local modulation of the potential energy inside a nanofluidic slit using simple geometrical indents. In thermodynamic equilibrium, following the Boltzmann distribution, the probability of finding a particle at a location $(x,y,z)$ depends exponentially on the potential energy at this location $U(x,y,z)$ \cite{israelachvili2011intermolecular}
\begin{equation}
    P(x,y,z)\propto \exp{\left[-\frac{U(x,y,z)}{k_\mathrm{B}T}\right]}, \label{eq:1boltzmann}
\end{equation}
with the Boltzmann constant $k_\mathrm{B}$ and temperature $T$. For charged nanoparticles, in the utmost vicinity of the solid walls, the contribution from dispersion interactions dominates. Further away, however, the potential energy is dominated by the electrostatic contribution, which is proportional to the electrostatic potential $\psi(x,y,z)$ \cite{israelachvili2011intermolecular}. Particles can be trapped in local minima of the potential energy. In nanofluidic slits with weakly overlapping electric double layers (EDLs) and the same wall charge polarity as that of the nanoparticles, recesses with depths comparable to the Debye length $\kappa^{-1}$ create local minima in electrostatic potential energy. The average escape time $t_\mathrm{esc}$ of a particle from a potential well of $\Delta U$ is then given by Kramer's formula \cite{Kramers1940} 
\begin{equation}
    t_\mathrm{esc}=t_\mathrm{r}\exp\left(-\frac{\Delta U}{k_\mathrm{B}T}\right), \label{eq:2kramer}
\end{equation}
where $t_\mathrm{r}$ is the characteristic diffusion time across the recess in absence of potential barriers. Given the exponential dependence, nanoscale objects can be trapped for extended periods of time ranging up to seconds \cite{Ruggeri2017}. 
Importantly, here, the potential energy landscape is essentially independent of the particle polarizability and can be tuned by the recess’ geometry, the nanofluidic gap, and by optimizing the electrolyte \textit{p}H to maximize the surface charge of the confining walls. 

The trapping of a variety of nanoscale objects down to single fluorophores has demonstrated the capabilities of this scheme \cite{Ruggeri2018a}. The subsequent monitoring of the average escape time of trapped entities has been introduced as a precise way to characterize charge at the single-molecule level \cite{Ruggeri2017}.
This enabled applied studies that, for example, determined the conformation of biomolecules \cite{Kloes.2022} or simultaneously characterized the charge and size of extracellular vesicles \cite{Hosseini.2021}.
Further applications of electrostatic fluidic traps include the orientation of anisotropic nanoparticles in shape \cite{Celebrano2012a} or charge \cite{Ruggeri2022} and their subsequent directed assembly \cite{Fringes.2019,Ruggeri2022}. Different implementations of the electrostatic fluidic trap have been reported,
among others the use of a nanopipette \cite{Kim2014}, a convex lens \cite{Berard.2014}, a semi-flexible membrane \cite{Hosseini.2021} or by leveraging elastic deformation in soft materials \cite{Gerspach2017}. Finally, the creation of more complex topographies, such as nanoscale rachets has recently enabled achieving directed particle transport \cite{Skaug2018} and sorting \cite{Nicollier2021} by using the principles of the electrostatic fluidic trap.

Despite its demonstrated ability to efficiently trap particles and these recent advancements, the electrostatic fluidic trap lacks tunability. It is inherently static, requiring the creation of specific device geometries for each experiment. Here, we introduce the use of gate electrodes embedded beneath the traps to dynamically control the wall-particle interactions by tuning the potential of the gate electrodes, and thus modulating the local surface charge.

The concept of using gate electrodes for field-effect control in microfluidics has long been used to improve the resolution of capillary electrophoresis \cite{Lee_1990,Ghowsi_1990,Ghowsi1991} or to control electroosmotic flow \cite{Schasfoort_1999,dehe2021,Schonecker2014,Dehe2020,Paratore2019}.
Gate electrodes have also been employed in sub-micrometer dimensions \cite{Hu_2012,Guan2012,Prakash2016}, e.g. for modulating ionic currents \cite{Gajar_1992} or manipulating water-dispersed nanoscale objects \cite{Karnik_2005,Fan_2005_PRL,Karnik_2006}. Modulations of surface potentials have been probed by observing the mobility of charged dye molecules \cite{Oh_2009}, and molecule-wall interactions were measured by infrared spectroscopy \cite{Oh_2008}.
In recent years, the gating of nanoscale geometries has mostly been driven by the nanopore community for control of ionic conductivity \cite{Nam_2009,Szleifer_2015,PerezMitta.2017} or translocation of molecules \cite{Hernandez_2015,Tsutsui_2021}, and even exploring novel transport mechanisms \cite{Ratschow2022}.

The fundamental idea of this work is to combine a defined wall topography with electrostatic gating to achieve controllable nanoparticle traps. 
We experimentally demonstrate achievable surface potential modulations of 50 mV at $p\mathrm{H}=6.6$, translating to modulations in the trapping potential energy of $0.7\,k_\mathrm{B}T$ for the given device geometry.
We use two distinct measuring schemes and formulate a parameter-free predictive model that quantitatively captures our experimental findings. 

 To this end, we fabricate a nanofluidic device by using a glass substrate and depositing a Cr-Pt-Cr electrode ($\SI{2}{nm}-\SI{6}{nm}-\SI{2}{nm}$) by electron beam evaporation. The electrode is patterned with a hexagonal lattice of circular holes 
 using optical lithography. This structure provides a wall topography of \SI{10}{nm} deep recesses with the gate electrode only present outside of the traps, while the bottom of the traps remains ungated. The electrode is followed by a layer of \SI{20}{nm} \ce{SiO2} deposited by atomic layer deposition (ALD), a \SI{500}{nm} layer of \ce{SiON} and a \SI{180}{nm} top layer of \ce{SiO2}, both deposited by plasma-enhanced chemical vapor deposition (PECVD). We choose the comparatively thick layer of \ce{SiON} for its high resistance to dielectric breakdown and symmetric resistance to both biases \cite{Paratore2019} and apply the \ce{SiO2} top layer for its well-known surface chemistry [Supporting Information (SI) \S1.1]. As the oxide deposition is nearly conformal, the topography imparted by the holes in the electrode is maintained over the entire stack of thickness $d=\SI{700}{nm}$ (Fig. 1a). Figs. 1b and c show the final surface topography of the top layer. The AFM image reveals traps with a hole diameter of $2r=\SI{1.1}{\micro m}$ and a depth of $\SI{10.8(0.6)}{nm}$. 
 Note that the bottom of the holes is electrode-free.  

\begin{figure}
  \includegraphics[width=\columnwidth]{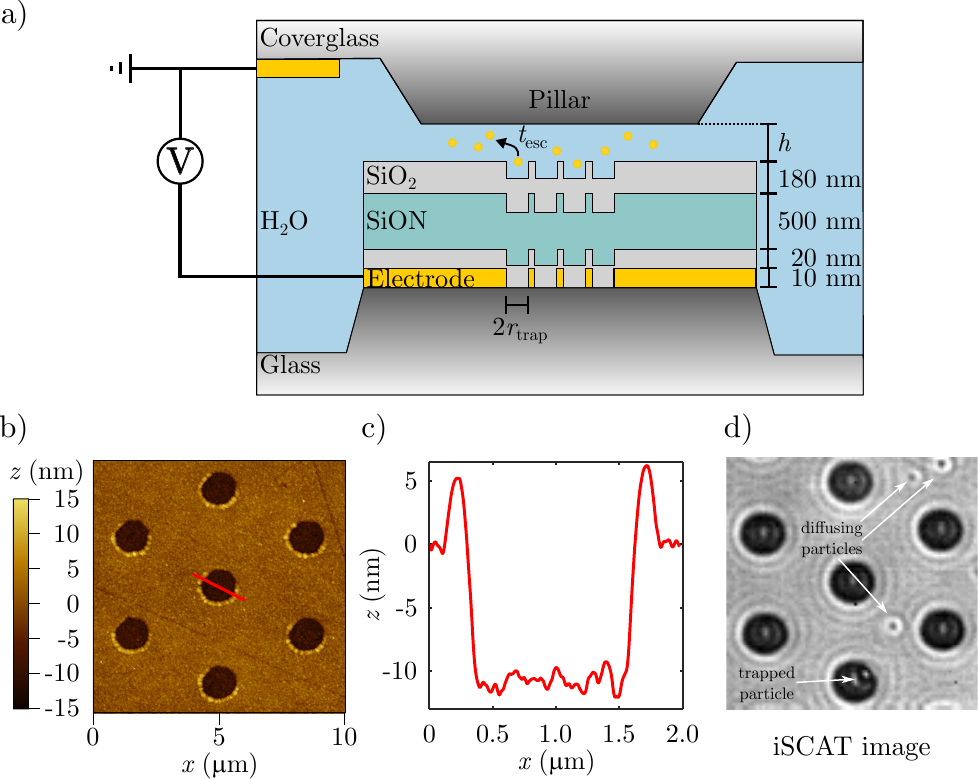}
  \caption{Experimental details. a) Schematic cross section of the device including the coverglass and nanoparticles. The topography imparted by the holes in the electrode in conformally mapped through all layers. b) AFM scan of the \ce{SiO2} top layer showing the hexagonal pattern of recesses. c) AFM trace across one single recess. Traps have a final diameter of \SI{1.1}{\micro m} and a depth of \SI{10.8}{nm}. d) Instantaneous interferometric scattering microscopy (iSCAT) image of diffusing and trapped nanoparticles on the same geometry as b). A corresponding movie can be found in the supporting information.}
\end{figure}

Overall, the nanofluidic device is a $15\times\SI{33}{mm}$ chip with an experimental area of $200\times\SI{200}{\micro m}$ (SI \S1.1, Fig. S2a-c). Before experiments, we ensure a sufficiently high dielectric breakdown resistance by placing a grounded electrolyte drop on the chip and applying step-wise increasing gate voltages to the electrode with time intervals of \SI{10}{s} and steps of \SI{50}{V} using a high voltage source measuring unit (Keithley 2410). The current is monitored to detect dielectric breakdown. We only accept a device if its leakage current is $\leq\SI{1}{\micro A}$ and if it is stable against breakdown up to gate voltages of $-\SI{300}{V}/+\SI{500}{V}$ (SI \S1.2). 

Throughout this work we use \SI{70}{nm} citrate-capped gold nanoparticles (Nanocomposix) due to their well-characterized zeta potential of \SI{-50(10)}{mV} at \textit{p}H values above $\approx4.5$ (SI \S1.3). We suspend the nanoparticles in DI water that becomes acidic due to dissociation and partial dissolution of the citrate, obtaining a \textit{p}H of 6.6 and a Debye length $\kappa^{-1}$ of \SI{13.3}{nm} (SI \S1.4). Nanoparticles interactions are negligible due to their low concentration. 
We place a drop of this nanoparticle suspension onto the device before creating a nanofluidic slit. The liquid is held in the slit by capillary forces. 

To achieve well-defined and controllable nanoscale slits, we employ the nanofluidic confinement apparatus, described under \cite{Fringes_Thesis,Fringes2018} and SI \S1.5. The fabricated nanofluidic device is attached to the apparatus and approached by a coverglass with a central glass pillar/mesa of \SI{100}{\micro m} lateral size and \SI{40}{\micro m} height. The pillar is etched into the glass using hydrofluoric acid, see Fig. 1a. The coverglass and the microscope objective are lowered by two linear piezo-stages (\SI{100}{\micro m}, Nano-OP100, Mad City Labs) which are attached to a coarse positioning stage (MT-84, Feinmess). This technique allows us to create nanofluidic slits down to the sub-\SI{100}{nm} range, bound by parallel surfaces with a tilt of less than \SI{1}{nm} per \SI{10}{\micro m} lateral distance, measured by interferometry between the two surfaces. For the experiments presented below, the slit height was $h=\SI{195\pm2}{nm}$. 
We image the nanoparticles by interferometric scattering detection (iSCAT) \cite{Lindfors2004,Fringes2016} (Fig. 1d), recording at 250 frames per second. 

To observe the effect of a voltage applied at gate electrode on the potential energy landscape experienced by diffusing nanoparticles, we track the particle centroids using TrackPy \cite{Trackpy}. We monitor the occupancy state of each trap over time as well as the position of each particle in the field of view. From this, we extract two statistical quantities: the escape time distribution of the confined particles \cite{Ruggeri2017} and the proportion of trapped particles. The escape time approach gives us relative changes from the reference situation with no applied gate voltage. Meanwhile, the proportion of trapped particles is directly linked to the Boltzmann relation of equation \ref{eq:1boltzmann} and provides absolute values of potential energy barriers $\Delta U$.
 
\begin{figure*}
  \includegraphics[width=0.6\textwidth]{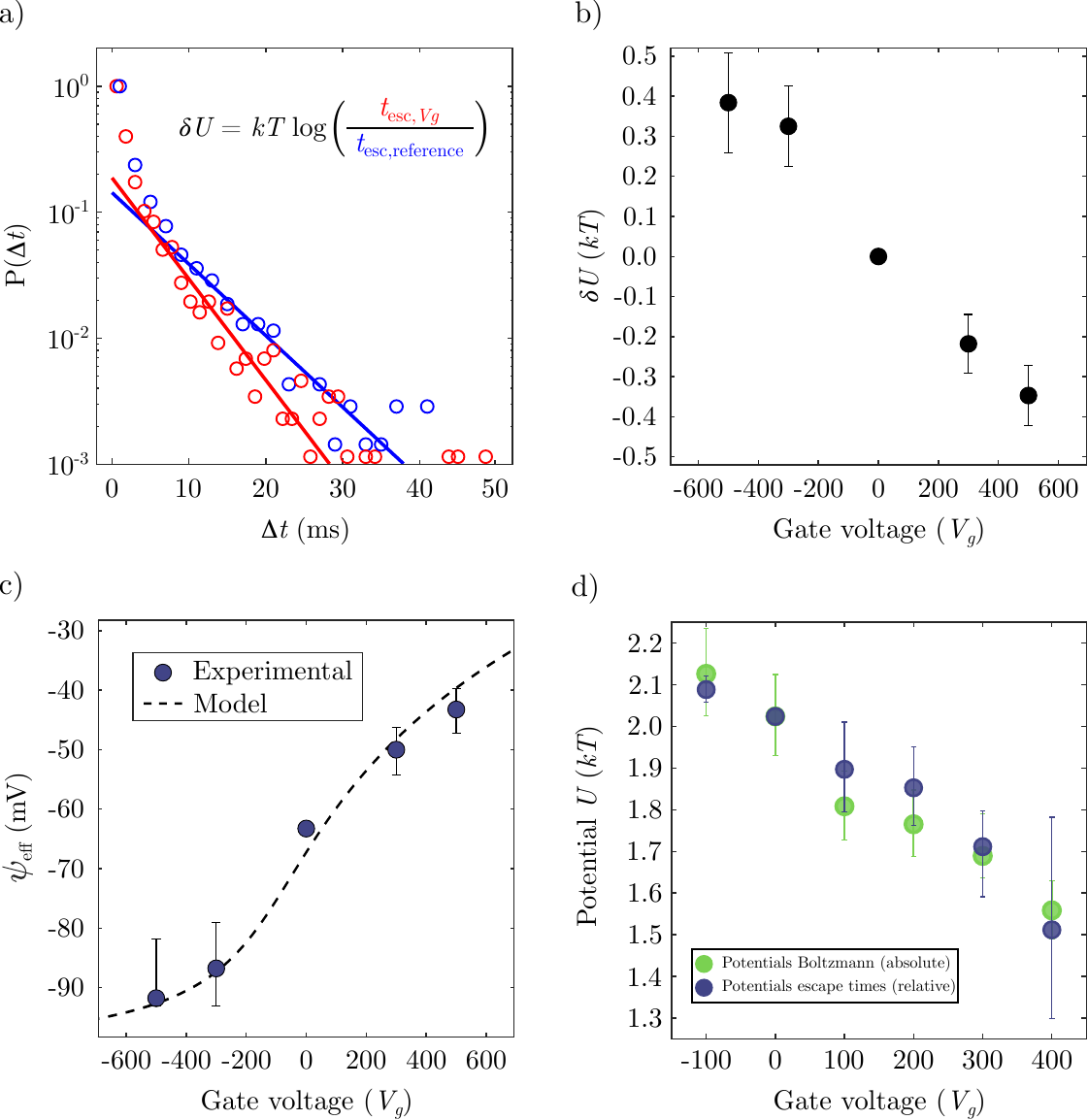} 
  \caption{Experimental characterization of the potential energy landscape experienced by nanoparticles. a) Comparison of the normalized probability distributions of escape times of nanoparticles $\mathrm{P}(\Delta t)$ with and without an applied gate voltage and extraction of the potential energy modulation $\delta U$. b) Potential energy modulations over gating field strength extracted from escape time measurements. c) Modulations in effective surface potential outside of the traps over gating field strength for the same data as in b) and predictions of the  parameter-free model of equations \ref{eq:4sigmachem}-\ref{eq:6psieff} (dashed line). In b) and c), each point represents one experiment as shown in a). Error bars are obtained from the standard error of the regression slope to the escape time distribution within each experiment. d) Comparison of the depth of the potential well over gate voltage for the escape time and Boltzmann measurements. Since the escape time measurements only provide relative changes, they are applied relative to the point at zero gate voltage. Error bars are obtained from the standard error of the regression slope to the escape time distribution for the blue data points, while we assume a systematic particle tracking error of 1\% to determine the error bars on the blue data points. }
\end{figure*}

Fig. 2a exemplarily shows the escape time distributions with and without an applied gate voltage $V_g = \SI{500}{V}$. The characteristic escape times $t_\mathrm{esc,i}$ are related to the slopes of the exponential fits (solid lines). Following Kramers relation, equation \ref{eq:2kramer}, the change in potential energy $\delta U=\Delta U_\mathrm{V_g}-\Delta U_\mathrm{reference}$ is given by $\delta U/k_\mathrm{B}T=\log\left(t_\mathrm{esc,V_g}/t_\mathrm{esc,reference}\right)$, see Fig. 2a. In this example, we see a clear decrease in escape time due to an applied positive bias voltage. The trap potential energy decreases by $0.35\,k_\mathrm{B}T$. We repeat the measurement for different gate voltages with reference measurements between each experiment, to avoid drifts due to possible induced changes in the surface chemistry. The results of the voltage sweeps are shown in Fig. 2b and c. Negative biases increase and positive biases decrease the potential well depth of the traps. 

To verify this result, we compare relative potential energy changes obtained from escape time measurements to absolute potential energy barriers $\Delta U$ obtained from the Boltzmann relation. Therefor, we compare the area-normalized ratio of trapped and freely diffusing particles and extract the potential energy for each applied voltage using equation \ref{eq:1boltzmann}. The good agreement between the two methods (Fig. 2d) corroborates our finding. Furthermore, it confirms that both methods are well suited for characterizing potential energy landscapes in electrostatic fluidic traps. Note that Fig. 2d shows a separate experiment from Figs. 2a - c, performed at a gap distance of \SI{163\pm 2}{nm} in a buffer of \SI{1}{mM} NaOH and \SI{2}{mM} acetic acid at a $p$H of 5.
The stronger confinement provided a trapping potential of 2.02 $k_\mathrm{B}T$ with no gate voltage vs. 1.44 $k_\mathrm{B}T$ for the experiment in DI water. Thus, observed potential modulations were proportionally stronger as compared to Fig. 2 b). 

The surface potential is the key parameter characterizing surface charge modulation. Hereunder, to convert the potential energies to changes in surface potential, we use EDL theory with the Derjaguin approximation \cite{Derjaguin1934} and the linear superposition approximation \cite{Bell_1970}. These are justified by the comparatively large nanoparticles and weak double layer overlap (Debye length $\kappa^{-1}=\SI{13.3}{nm}$, gap distance $h=\SI{195\pm 2}{nm}$, nanoparticle size $d_\mathrm{NP}=\SI{70}{nm}$, SI \S2.1 and \S2.2).
Fig. 2c shows the changes of the depth of the potential well converted to effective surface potentials. 

\begin{figure}
  \includegraphics[width=0.9\columnwidth]{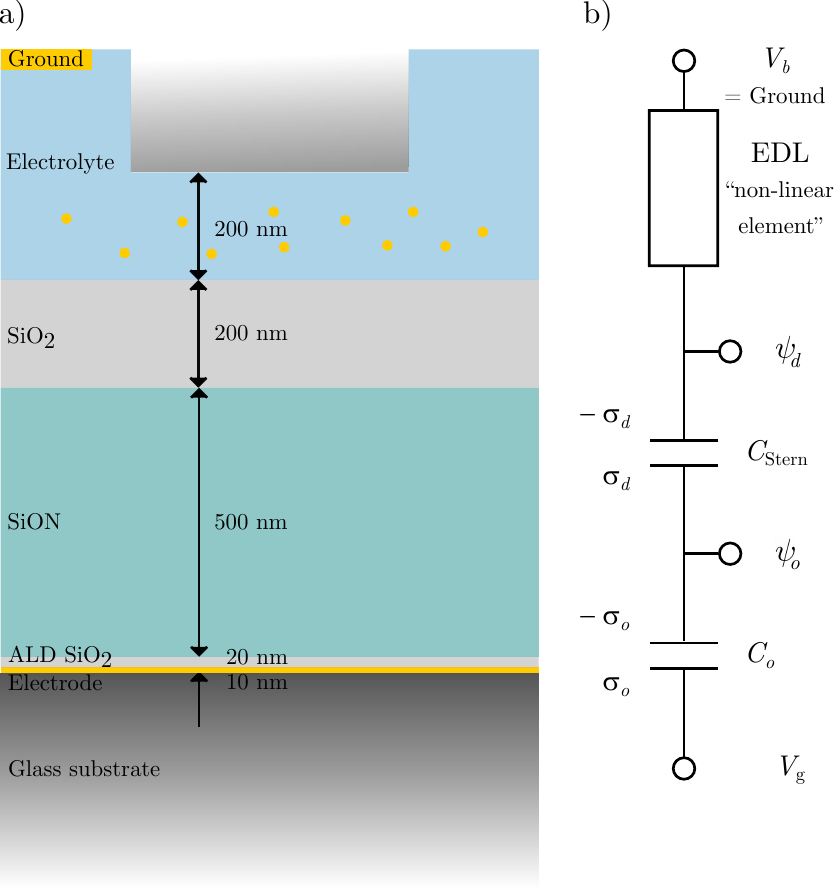}
  \caption{The dielectric stack. a) Schematic cross section of the dielectric stack and the nanofluidic slit drawn to scale. b) Equivalent circuit model of the dielectric stack shown in a), with the EDL represented by a non-linear element.}
\end{figure}

Generally, the change of effective surface potential of the gated wall is a combination of the field effect and changes in the surface charge density. Both are coupled through the EDL structure and the chemical equilibrium of reactive surface groups \cite{Behrens2001}. At \textit{p}H values above the point of zero charge, the surface chemistry of \ce{SiO2} in water is governed by the deprotonation of silanol groups \cite{vandenBerg2006,Ratschow.2023} 
\begin{equation}
    \ce{SiOH}\Longleftrightarrow \ce{SiO-} +\ce{H+}.
\end{equation}
To understand the complex interplay of EDL structure, surface chemistry, and gating, we resort to a model reported by Jiang and Stein \cite{Jiang2010} and adapt it to the present case. Due to the weak diffuse layer overlap, the gated wall is --~to a good approximation~-- in contact with a grounded liquid reservoir. The device can then be represented by an equivalent circuit comprising the electrode of potential $V_g$, the dielectric stack capacitance $C_0\approx\SI{5e-5}{F/m^2}$, the Stern layer capacitance $C_\mathrm{Stern}=\SI{2.9}{F/m^2}$ \cite{Hiemstra1989}, and a non-linear element that includes the EDL and the surface chemistry (Fig. 3b, SI \S2.4). The surface chemistry equilibrium yields an expression for the surface charge density
\begin{equation}
    \sigma=\frac{-e\Gamma}{1+10^{(p\mathrm{K}-p\mathrm{H})}\exp\left(-\frac{1}{k_\mathrm{B}T}\frac{\sigma+C_0V_g+C_\mathrm{Stern}\psi_d}{C_\mathrm{Stern}+C_0}\right)},\label{eq:4sigmachem}
\end{equation}
with the elementary charge $e$, silanol group density $\Gamma=\SI{8}{nm^{-2}}$ \cite{Behrens2001}, dissociation constant $p\mathrm{K}=7.5$ \cite{Hiemstra1989,Behrens2001}, and diffuse layer potential $\psi_d$. The surface charge density is also related to $V_g$ through Gauss' law. For the present case, it reads (SI \S2.5)
\begin{equation}
    \sigma=\frac{C_0+C_\mathrm{Stern}}{C_\mathrm{Stern}}\frac{2\varepsilon_0\varepsilon_r\kappa k_\mathrm{B}T}{e}\sinh\left(\frac{e\psi_d}{2k_\mathrm{B}T}\right)+C_0(\psi_d-V_g), \label{eq:5sigmagauss}
\end{equation}
 with the vacuum and relative liquid permittivity $\varepsilon_0$ and $\varepsilon_r$. The diffuse layer potential $\psi_d$ itself is not experimentally accessible but can be converted to the effective surface potential $\psi_\mathrm{eff}$ by \cite{Jiang2010}
 \begin{equation}
     \psi_\mathrm{eff}e/k_\mathrm{B}T=4\tanh(\psi_de/k_\mathrm{B}T). \label{eq:6psieff}
 \end{equation}
Equations \ref{eq:4sigmachem}-\ref{eq:6psieff} are a model for the effective surface potential above the gate electrode. It incorporates the Gouy-Chapman analytical solution of the one-dimensional Poisson-Boltzmann equation for an extended flat plate. Because of the low aspect ratio of the recesses, the model is applicable to the present problem under the long-wavelength approximation \cite{Leal.2012}. 
  
  \begin{figure}
  \includegraphics[width=0.9\columnwidth]{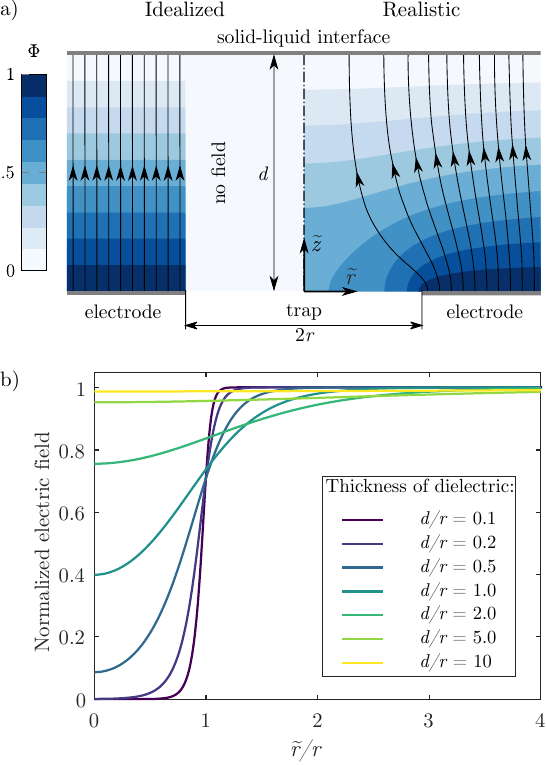}
  \caption{Electrostatic fringing. a) Comparison of the idealized situation without fring fields (left) and simulations of the realistic electrostatic potential and field in the dielectric stack (right). Electric fields leak into the traps and gating occurs both inside and out of the recesses. b) Gating field strength at the solid-liquid interface over trap radius for different aspect ratios $d/r$. The trap spans $\Tilde{r}/r=0$ to 1. While large traps in thin substrates ($d/r<1$ come close to the idealized situation, fringing obstructs selective gating outside the traps for small traps in thin substrates ($d/r>1$).}
\end{figure}
 
 To predict the modulated surface potentials with the model, we need to account for electrostatic fringe fields within the dielectric stack of thickness $d$. With the bottom of the holes of radius $r$ being electrode-free, for a sufficiently wide trap ($d/r\ll1$) the gating would only affect the areas outside of the traps. However, the smaller the trap, the more the electric field leaks into the trap, thereby reducing the expected modulation of the surface potential. Fig. 4a compares the idealized and realistic field distributions in the dielectric stack. 
 Since the electric field density relates to the the field effect acting on the solid-liquid interface, it is clear that in the realistic scenario there is gating also within the trap and, simultaneously, the gating outside is weaker than in the idealized case. 
 We quantify the effect through finite-element simulations of the electrostatic potential in the dielectric stack (details SI \S2.6). Fig. 4b shows the normal electric field at the solid-liquid interface scaled by the expected, idealized value $V_g/d$ (Fig. 4a). While for large traps $d/r\ll1$ the gating effect is close to the ideal situation, for small traps $d/r\gg1$ selective gating outside of the traps becomes impossible. For the present case of $d/r=1.3$, we calculate the average normalized electric field at the solid-liquid-interface, measuring the gating strength, inside and outside the trap as $0.67$ and $0.9$, respectively. Even though the gating substantially influences both regions, it is stronger outside of the traps and allows for deliberate modulations of the potential well depth.
 
We correct $V_b$ by using these geometrically determined values for the effectiveness of gating as prefactors and solve equations \ref{eq:4sigmachem}-\ref{eq:6psieff} both within and outside of the traps. These equations constitute a predictive, parameter-free model for the effects of gating on electrostatic fluidic traps. Fig. 2c demonstrates excellent agreement between the model (dashed) and our experiments based on escape time measurements. 

Overall, we achieve modulations in effective wall potential (equation \ref{eq:6psieff}) between \SI{-42}{mV} and \SI{-91}{mV} and respective changes in potential energy by $0.7\,k_\mathrm{B}T$ for gate potentials between $-500$ to \SI{500}{V} (Figs. 2b and c). The measurement schemes based on equilibrium (Boltzmann) and non-equilibrium (escape time) processes yield comparable results when characterizing gated electrostatic fluidic traps. For the present trap and stack geometry, the gating is less effective than anticipated due to fringe fields. The results in Fig. 4b suggest that wider traps or thinner dielectrics would improve the modulation capability. 
Generally, the effects of fringe fields and dielectric breakdown impose physical restrictions on the gating. There is a trade-off between modulation capability and precise localization. 
Decreasing $d/r$ by decreasing the thickness of the dielectric stack would reduce fringing but it would lead to dielectric breakdown at lower applied gate voltages. Thus it would decrease the achievable modulation. On the other hand, larger traps reduce fringing and allow for higher modulations in potential energy at the cost of less precise lateral localization of nanoparticles. In addition to these considerations, the main bottleneck seems to lie at the materials level. We propose that future work should focus on optimizing the dielectric properties to improve the resistance to breakdown and achieve stable and reproducible devices at lower stack thicknesses or higher gate voltages. 

To conclude, we have demonstrated that gate electrodes enable systematic control of electrostatic fluidic traps. Control is achieved by selectively modulating the effective surface potential of the confining wall outside of the geometrically induced traps, while leaving the inside of the traps largely unaffected. Although electrostatic fringe fields in the dielectric stack can mitigate the selective gating, we achieved modulations in the potential energy of $0.7\,k_\mathrm{B}T$ and respective changes in effective wall potential of $\approx\SI{50}{mV}$. This study thus provides a proof-of-concept for gate-modulated traps. We have introduced a parameter-free predictive model for the gating that agrees quantitatively with our experiments. 
The model could be used to evaluate other device designs -- regarding material composition, surface charge density or trap geometry -- to optimize trap modulation. 
Gated electrostatic fluidic traps enable real-time control and switching of nanofluidic potential energy landscapes experienced by nanoparticles.
More fundamentally, our results suggest that controllable nanoparticle traps are also feasible without an imparted surface topography -- by patterned electrodes under flat walls. This could allow for deliberate activation and even complete deactivation of trapping effects. Finally, modulating the effective surface potential in nanofluidic confinement would permit deliberate shifting of DLVO curves, which could prove useful for guided assembly \cite{Holzner2011a}.
 
\begin{acknowledgements}

The authors thank Maximilian T. Sch\"ur, Lucio Isa, Jan Eijkel, Derek Stein and Philippe Renaud for helpful discussions. 
 
A.W.K. proposed and supervised the work, F.P. and P.M.N. conceived the experiment, P.M.N., F.R., U.D., and F.P fabricated the devices, P.M.N. conducted, and evaluated the experiments, P.M.N. and A.D.R. developed the model, A.D.R. performed the simulations, P.M.N., A.D.R., S.H., and A.W.K. contributed to the interpretation of the results, A.D.R. wrote the manuscript with input from all authors, P.M.N. and A.D.R. wrote the Supporting Information.

This work was supported by the Swiss National Science
Foundation, grant SNSF No. 200021-179148 (P. M. Nicollier, F. Ruggeri, and A. W. Knoll). 

\end{acknowledgements}

\bibliographystyle{apsrev4-1} 
\bibliography{main.bib}  

\clearpage


\section*{Supplementary material}

# Gate Electrodes Enable Tunable Nanofluidic Particle Traps
## – Supporting Information –

Philippe M. Nicollier,[1] Aaron D. Ratschow,[2] Francesca Ruggeri,[1] Ute Drechsler,[1] Steffen Hardt,[2] Federico Paratore,[1,3,*] and Armin W. Knoll[1,†]

[1] *IBM Research Europe - Zurich, Säumerstrasse 4, CH-8803 Rüschlikon, Switzerland*
[2] *Institute for Nano- and Microfluidics, TU Darmstadt, Alarich-Weiss-Strasse 10, D-64287 Darmstadt, Germany*
[3] *Laboratory for Soft Materials and Interfaces, Department of Materials, ETH Zürich, Vladimir-Prelog-Weg 5, CH-8093 Zürich, Switzerland*

(Dated: September 4, 2023)

The Supporting Information consists of two main sections. §1 provides details about the manufacturing of the devices and the experimental materials and procedures. In §2, details about the analytical model and its derivation are given.

## Contents

## §1 Experimental Details

### §1.1 Device Structure

The nanofluidic device for the present study should exhibit (i) simple & reproducible fabrication, (ii) good resistance to dielectric breakdown, (iii) a top surface consisting of $SiO_2$ to have a known surface chemistry, and (iv) compatibility with the nanofluidic confinement apparatus (§1.5) to create stable, nanofluidic gaps. The first two points were investigated by Paratore et al. [S1]. They focused on dielectrics fabricated by plasma-enhanced chemical vapor deposition (PECVD) with some inter-layers fabricated by atomic layer deposition (ALD). Regarding the breakdown characteristics, they compared a range of different dielectric stacks and quantified their breakdown characteristics. A summary of their measurements is shown in Fig. S1.

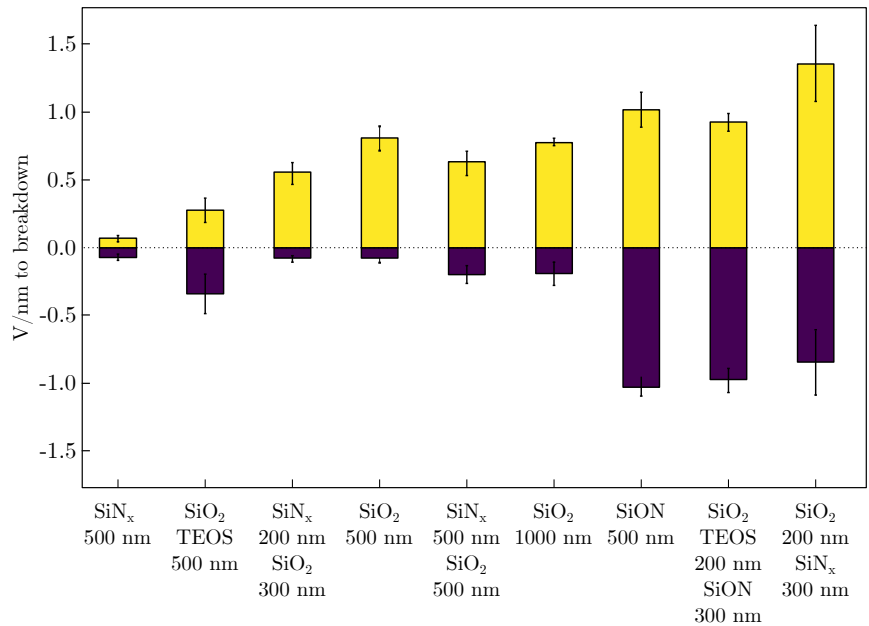

FIG. S1. Different dielectric stacks tested with regard to their breakdown characteristics. Data and figure adapted from Paratore *et al.* [S1]. A combination of 500 nm SiON with an additional layer of 200 nm was used due to its symmetric breakdown characteristics.

A crucial element is the dielectric breakdown resistance at both bias polarities. SiON displays the highest symmetry in this case. We thus used a stack composed of a 500 nm thick layer of SiON covered by a thinner layer of $SiO_2$. In the fabrication process, we also noticed that a first, thin ALD layer of $SiO_2$ or $HfO_2$ significantly improved the reproducibility of our devices.

A cross section of the final device design is shown in Fig. S2a. The glass substrate was chosen for its insulating properties. A Cr-Pt-Cr electrode (typical thickness: 2 nm − 6 nm − 2nm) was deposited by electron beam evaporation and covered with 20 nm of $SiO_2$ deposited by atomic layer deposition. Then, a 500 nm layer of SiON was deposited by PECVD, followed by a 180 nm top layer of $SiO_2$. Fig. S2b schematically shows the geometry of a single chip. It has dimensions of 15 x 33 mm. The roughly 0.5 $cm^2$ large contact pad narrows down into a 20 $\mu$m wide and 2 cm long electrode strip which connects the contact pad to the experimental area of the device.

The ridge indicated in Fig. S2 b) is a 200 $\mu$m wide region around this electrode strip, that has been elevated by 50 $\mu$m using a concentrated HF etch process. It enables an unobstructed approach between the device surface and the pillar on the coverslip (§1.5).

The 'experimental area' is a 200 x 200 $\mu$m square where we conducted all of the reported experiments. At this location, we approached the pillar on the coverslip to effectively create the nanofluidic confinement. The experimental area consists of a patterned region of the electrode achieved by optical lithography and a lift-off process. This is done at the same time as the electrode is defined and before depositing the dielectric. We patterned an array of holes arranged in an hexagonal grid as well as a checkerboard pattern and thin channels. Only the hole array is used in the subsequent experiments.


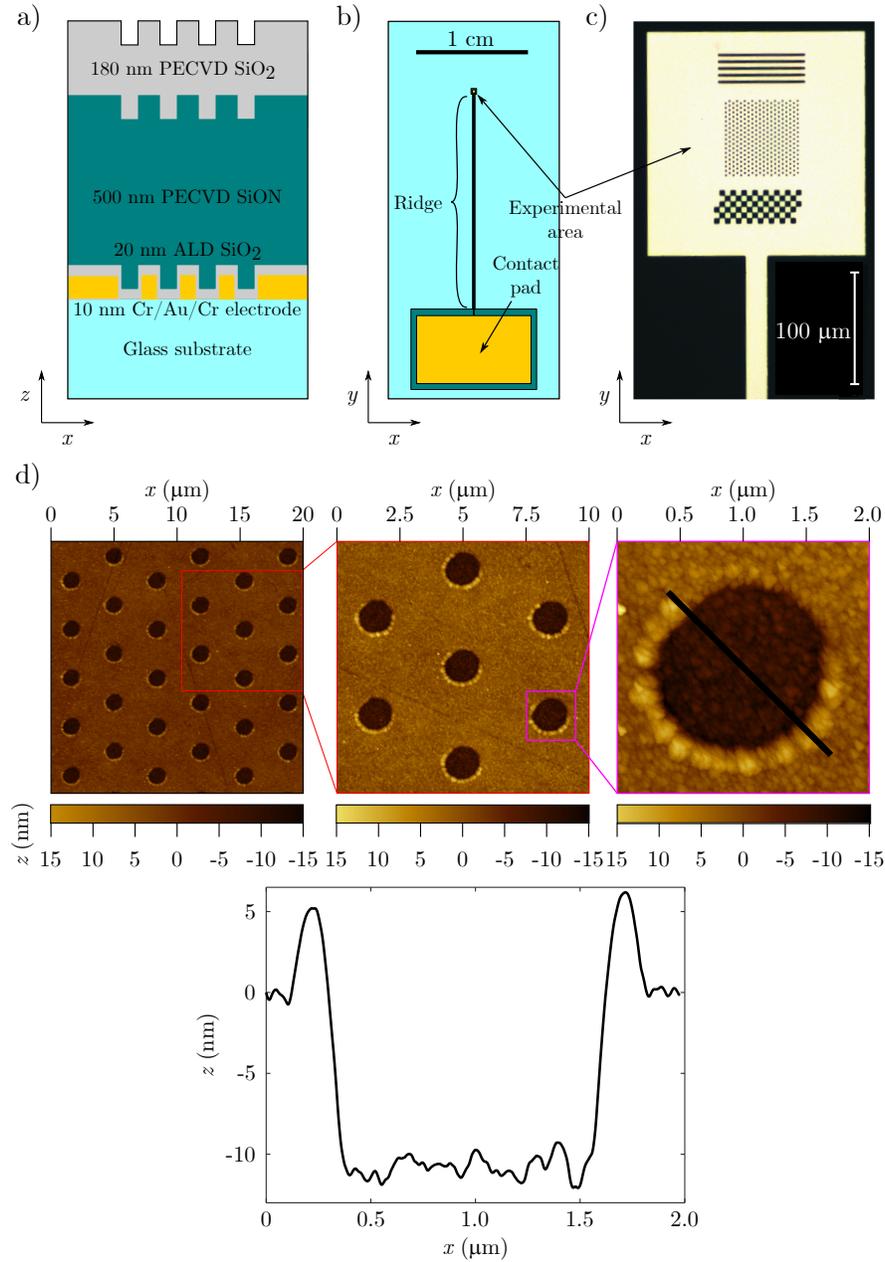

FIG. S2. Device structure and AFM scans of the experimental area. a) The layer composition of the device: 10 nm Cr-Au-Cr electrodes are coated with 20 nm ALD $SiO_2$, followed by a 500 nm layer of SiON and 180 nm $SiO_2$, both deposited by PECVD. b) The chips have an area of 33 x 15 mm. c) Optical micropscopy image of the testing area of the device: the electrode contains holes of different geometries. The reported experiments were all performed in the central trap array. d) AFM traces of the trap array: the dielectrics are deposited conformally, the topography is transferred from the electrode layer to the top layer. Holes have a final diameter of 1.1 $\mu$m, a depth of 10.8 nm, and present 5-6 nm ridges around the edge due to the lift-off process used to structure the electrodes.

## §1.2 Dielectric Breakdown

Before experiments, we applied gate voltages between the gate electrode and a drop of electrolyte placed on top of the experimental area of the device to ensure a sufficiently high breakdown resistance for subsequent experiments. The drop was in contact with a grounded electrode, while the voltage was applied to the pad connecting the gate electrode. Typically, we increased the bias in steps of 50 V held for 10 s, using a high-voltage source measuring unit (Keithley 2410) controlled by either a Matlab or a Labview script. Fig. S3 illustrates the voltage step function and

the measured current for a typical device. For this particular sweep, a sharp increase in current is measured as the gate voltage is switched to $V_g = 450\,\text{V}$, indicating dielectric breakdown of the oxide. Devices for experiments were only used with leakage currents $\leq 1\,\mu\text{A}$ and were required to be stable against breakdown up to gate voltages of - 300 V/+ 500 V. The resistance to negative biases was slightly less than for positive biases.

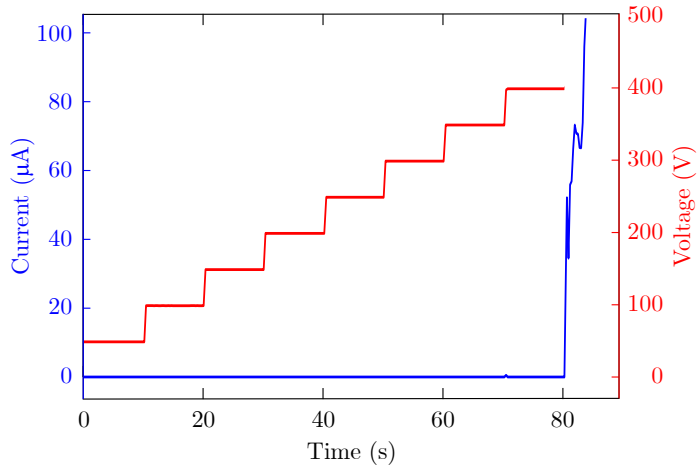

FIG. S3. Procedure to test the dielectric breakdown of the devices: the gate voltage is increased in 50 V/10 second intervals, until a sharp increase of the current is observable – this corresponds to the dielectric breakdown of the oxide layer.

## §1.3 Particles

Particles used in this work were 70 nm citrate-capped gold nanoparticles purchased from Nanocomposix. As the point of zero charge is below a $p$H of 2 for citrate and at a $p$H of 3.8 for silica surfaces, most practical experiments in nanofluidic confinement with glass surfaces have to be performed at $p$H values above $\approx 4.5$ to ensure strong enough electrostatic repulsion and avoid particle adsorption on the confining walls. The data displayed in Fig. S4 indicates that the $\zeta$-potential of citrate-capped gold nanoparticles in such conditions is of $50 \pm 10$ mV.

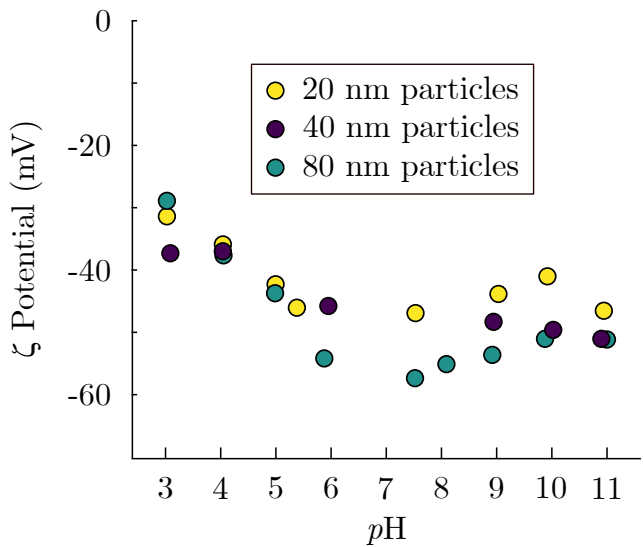

FIG. S4. Zeta potential data provided by the manufacturer of the Nanocomposix citrate-capped gold nanoparticles as a function of $p$H. Retrieved and adapted from www.nanocomposix.com. Last website visit: 01.12.2022.

### §1.4 Determination of Salt Concentration

To characterize the particle dispersions used in nanofluidic experiments, it is useful to know both $p$H and the salt concentration of the solution. They are required to estimate the zeta potential of the confining walls and to calculate the Debye screening length, which gives us the length scale of electrostatic interactions. We use a compact water quality meter, the LAQUAtwin-EC-33 from Horiba Scientific, to measure the conductivity of the dispersions, and the LAQUAtwin-pH-33 to measure their $p$H. We estimate the Debye length from the conductivity measurements with Kohlrausch's law [S2], which states that the molar conductance of a strong electrolyte $\sigma$ is:

$$\sigma = \sigma_0 - S\sqrt{c}, \tag{S1}$$

where $\sigma_0$ represents the limiting molar conductance, $S$ is a constant and $c$ is the molar concentration of the electrolyte. The conductivity measured by the conductivity meter is $\gamma$ and can be approximated by

$$\gamma = \sigma \cdot c\,[\mathrm{S/m}] \approx \sigma_0 \cdot c. \tag{S2}$$

For low concentrations, the conductivity depends linearly on the concentration of the ionic species. Since the conductivity of sodium citrate solutions is very similar to sodium chloride solutions [S2], we thus employ a calibration curve measured for NaCl solutions [S3], as shown in Fig. S5.

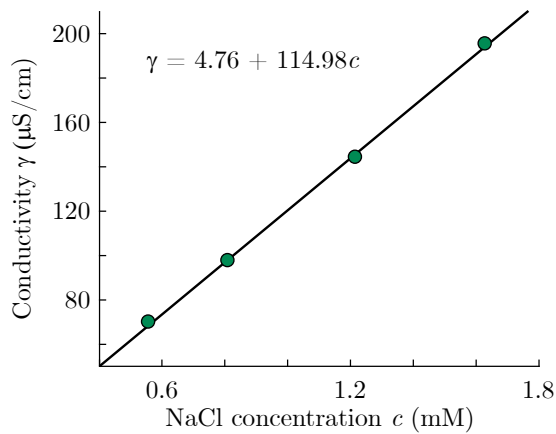

FIG. S5. Calibration curve used to deduce NaCl concentrations from conductivity measurements. Data adapted from Ruggeri *et al.* [S3]

The slope of 115 $\mu$m/(cm·mM) is in good agreement with the theoretical prediction, considering the limiting conductivities of Na$^+$ [S4] and Cl$^-$ [S5] which predict a specific conductance of 126.6 $\mu$m/(cm·mM) [S2]. Even though the citrate ion is trivalent, its counter-ion Na$^+$ is the determining species with respect to the double layer structure around the particles, as we assume that citrate molecules are preferentially adsorbed to the particle surface.

### §1.5 The Nanofluidic Confinement Apparatus

The nanofluidic confinement apparatus allows one to create nanometer gaps between the sample stage and a coverglass and control their separation with nanometer accuracy. The parallel alignment of both surfaces can be controlled down to 1 nm per 10 $\mu$m lateral distance. The gap can be dynamically adjusted, short-circuiting usually fabrication-intensive nanofluidic experiments. Moreover, it allows a tunable gap, enabling in situ changes of the experimental parameters. A schematic of the experimental set-up is shown in Fig. S6.

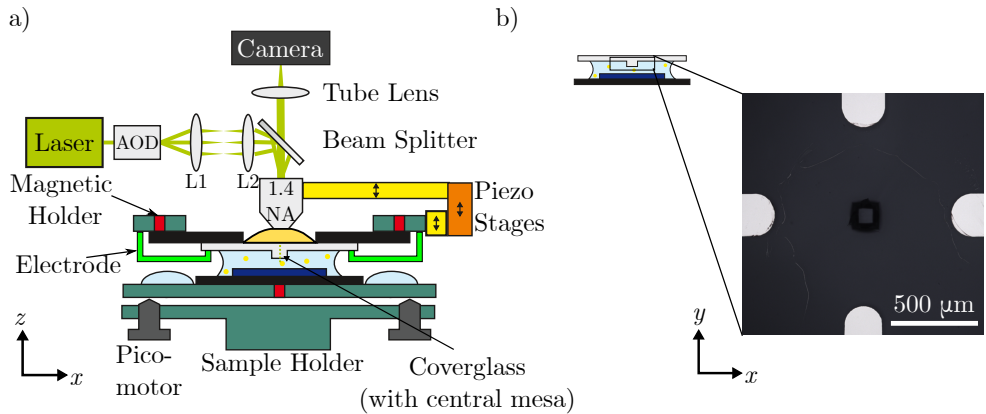

FIG. S6. a) Cross section of the experimental set-up, adapted from Fringes *et al.* [S2, S6]. b) Both optical and geometrical access is improved by using a coverglass with a central mesa. A microscope image is shown in the inset. The mesa is around 50 µm high and 200 µm wide. The electrodes surrounding the pillar can also be seen.

The nanofluidic gap is created between a sample with a nano-patterned surface and a coverglass. The sample is glued with wax onto a steel sheet which is held in place on a magnetic sample holder. The magnets are drawn with red color in Fig. S6a. The coverglass is glued onto a steel plate which is held upside down by magnetic holders. The center of the coverglass is shown in Fig. S6b: a central glass pillar/mesa of 100 µm lateral size and 40 µm height is etched into the glass using hydrofluoric acid. This is needed to allow an undisturbed approach to the region of interest, as well as to ensure a good optical access. The glass pillar is surrounded by four gold electrodes arranged in a square pattern, which are evaporated around the central pillar. The vertical and horizontal distance between the edges of two opposing electrodes is 1 mm. These electrodes are connected via copper tape and copper clamps to ground. Due to the isotropic etch performed by the HF acid, the side-walls of the pillar are not straight but have a 45° angle. This explains the apparent shadow around the pillar in Fig. S6b. The electrodes are placed in the region around the pillar, which can be seen as a 'microfluidic' reservoir around the 2D nanofluidic slit formed between glass pillar and substrate.

To create the nanofluidic gap, the top half of the set-up, which includes the microscope objective and the coverslip/coverslip holder, is lowered towards the substrate. In this regard, two linear piezo-stages (100 µm, Nano-OP100, Mad City Labs) are used, which are attached to a coarse-positioning stage (MT-84, Feinmess). To enable the close approach between the two surfaces and to be able to create small nanofluidic slits in the sub-100 nm range, the parallel alignment of the surfaces is crucial. The interferometric signal we measure from the reflected light can be employed to align the surfaces of the pillar and the bottom surface: we monitor the signal intensity as we oscillate the coverslip distance to the substrate. Subsequently, the intensity at the four edges of the field of view is measured.

Finally, it is also important that the entire system is parallel with respect to the table plane. This is needed to ensure being able to move in the $x-y$ direction even when the two confining surfaces are close to one another. This is ensured by three manual screws mounted on the coverslip holder, which allow to tilt the coverslip with respect to the table. This is optimized by translating the sample in the $x-y$ plane and observing the appearance of fringes after the surface has been parallelized with respect to the pillar plane.

## §1.6 Optical Set-up

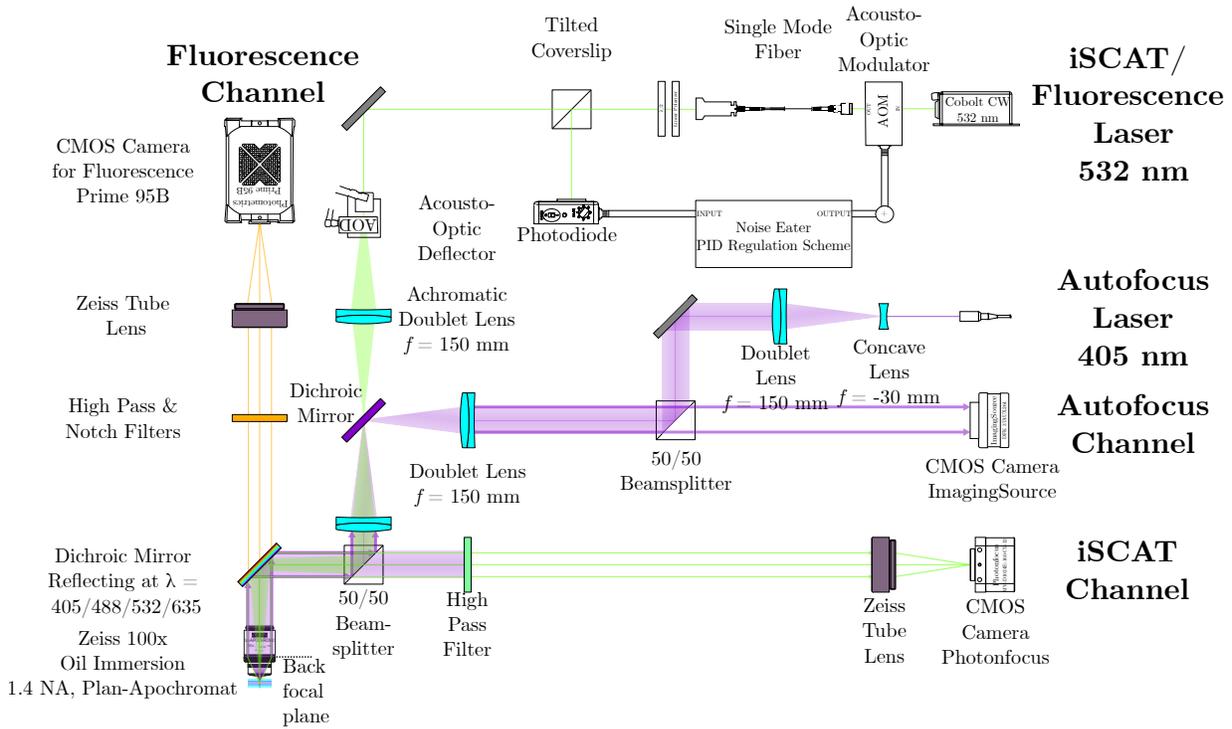

FIG. S7. Schematic drawing of the main components of the imaging set-up. The beams propagating towards the objective are drawn as solid lines/expanded lines while the rays propagating towards the camera are drawn with arrows. A detailed description is given below.

The optical set-up is used both for imaging particles confined in the nanofluidic slit, as well as for monitoring the gap distance between the confining walls. There are two lasers used and three detection channels in the optical set-up:

- The iSCAT channel for imaging particles using interferometric scattering microscopy, as well as to monitor the gap distance by the reflected intensity. A 532 nm CW laser (Cobolt Samba 50 mW) is used as a light source.

- The fluorescence channel for imaging fluorescent emitters was not used in this work.

- The focus channel: used to determine the focus position of the imaging laser source by measuring the objective-coverslip distance.

## §2 Modelling

### §2.1 The Linear Superposition Approximation

When distances between interacting surfaces are larger than the Debye length, the so-called linear superposition approximation can be applied [S7]: potentials from both confining walls can simply be added. For similarly charged surfaces, a potential extremum will thus be located at the midplane. Correspondingly, the counter-ion concentration will be lowest at this location. If the gap is asymmetric, the minimum plane will be shifted towards the surface of lower charge.

The linear superposition approximation is valid for separations above $\kappa^{-1}$. At closer distance, the potential distributions of the surfaces are coupled. Additionally, charge regulation phenomena come into play [S8].

### §2.2 The Derjaguin Approximation

While the interaction between two planar surfaces can be described by applying the superposition principle, the situation is more difficult when we deal with nonplanar geometries. In this case, the Derjaguin approximation [S9] is useful. It relates the force between two convex shapes situated at a distance $h$ from one another to the corresponding interaction $W(h)$ between two infinite parallel plates at the same distance:

$$F(h) = 2\pi R_{\text{eff}} W(h) \tag{S3}$$

where $R_{\text{eff}}$ defines an effective radius depending on the geometry of the two interacting surfaces. For two interacting spheres of radii $R_1$ and $R_2$, it can be shown by integrating the force between infinitesimal rings at the surface of the two spheres, that equation S3 becomes

$$F(h)_{\text{sphere-sphere}} = 2\pi \left( \frac{R_1 R_2}{R_1 + R_2} \right) W(h)_{\text{plane-plane}}. \tag{S4}$$

From this expression, the force between a sphere and a plane can be extracted by setting $R_2 = \infty$, which yields:

$$F(h)_{\text{sphere-sphere}} = 2\pi R_1 W(h)_{\text{plane-plane}}. \tag{S5}$$

Both the Derjaguin and the linear superposition approximations [S7] are used in this work to calculate the potential energies experienced by particles confined to nanofluidic, patterned slits. In our experiments we are always in a range where $\kappa h > 1$ and $\kappa r > 2$. Applying these approximations thus leads to negligible errors.

### §2.3 The Electrostatic Fluidic Trap

Consider the case of a nanofluidic gap of height $h$ with one of the confining walls patterned with a surface topography of depth $d(x, y)$, as shown in Fig. S8. A spherical nanoparticle of radius $r$ is situated at a vertical distance $z_p$ from the lower confining surface. This situation is sketched in Fig. S8a. We assume that the surface charge on both particle and confining surfaces has the same sign and that the linear superposition approximation is valid.

Applying the Derjaguin and linear superposition approximations, the interaction potential energy is a sum of two particle-plane interactions. With the effective surface potentials of the sphere, $\psi_S$, and the two planes, $\psi_{P,1}$ and $\psi_{P,2}$, the interaction potential energy of a particle corresponding to the situation illustrated in Fig. S8a can be expressed as

$$W(x, y, z_p) = W_0 r \psi_S \left( \psi_{P,1} e^{-\kappa(z_p - r)} + \psi_{P,2} e^{-\kappa(h + d(x,y) - z_p - r)} \right), \tag{S6}$$

with $W_0 = 4\pi\epsilon_0\epsilon_{H_2O}$. For equal surface potentials, $\psi_S = \psi_{P,1} = \psi_{P,2}$, the mid-plane interaction potential energy at the location of the recess will be decreased by a value

$$\Delta W = W_0 r \psi_S^2 \cdot e^{\kappa d} \tag{S7}$$

compared to the slit in the absence of a geometrical recess. The Boltzmann relation states that the probability of finding a particle at a location $(x, y, z)$ is proportional to the exponential of the potential energy at this location $W(x, y, z)$:

$$P(x, y, z) \propto \exp\left( -\frac{W(x, y, z)}{kT} \right). \tag{S8}$$

The probability of finding the particle at the recess location thus scales exponentially with the potential difference induced by the recess.

An entropic factor contributes further to the occupation probability of the trap. To calculate the total potential energy of a particle in an electrostatic fluidic trap, one needs to take into account the positional entropy of the particle along the z-axis [S10]. We therefore integrate the three-dimensional probability density

$$P(x, y) = C \int_{z=r}^{z=d(x,y)+h-r} e^{-W(x,y,z)/kT} \mathrm{d}z, \tag{S9}$$

where $C$ is a normalization constant. The potential energy $U(x, y)$ up to a reference potential $U_0$ is given by

$$U(x, y) = -\log(P(x, y)) + U_0. \tag{S10}$$

One can estimate the average escape time for a particle located in the potential energy well by using Kramer's formula [S11]:

$$\tau \propto \exp\left(\frac{e\Delta U}{kT}\right). \tag{S11}$$

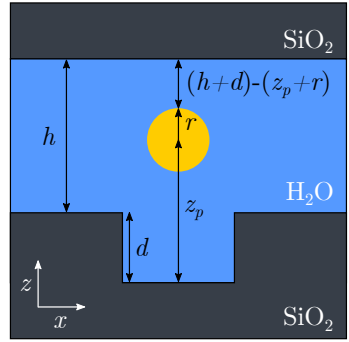

FIG. S8. Schematic of a gold nanoparticle in a patterned nanofluidic gap.

### §2.4 Equivalent Circuit and Capacitances

When gating a nanofluidic device, the resulting effect on the electric double layer will be a combination of the field effect and of changes in the surface charge density of chemically reactive surface groups. The theoretical modeling of this electric modulation for flat walls has been described by Jiang and Stein [S12] and will provide the basis to rationalize our results.

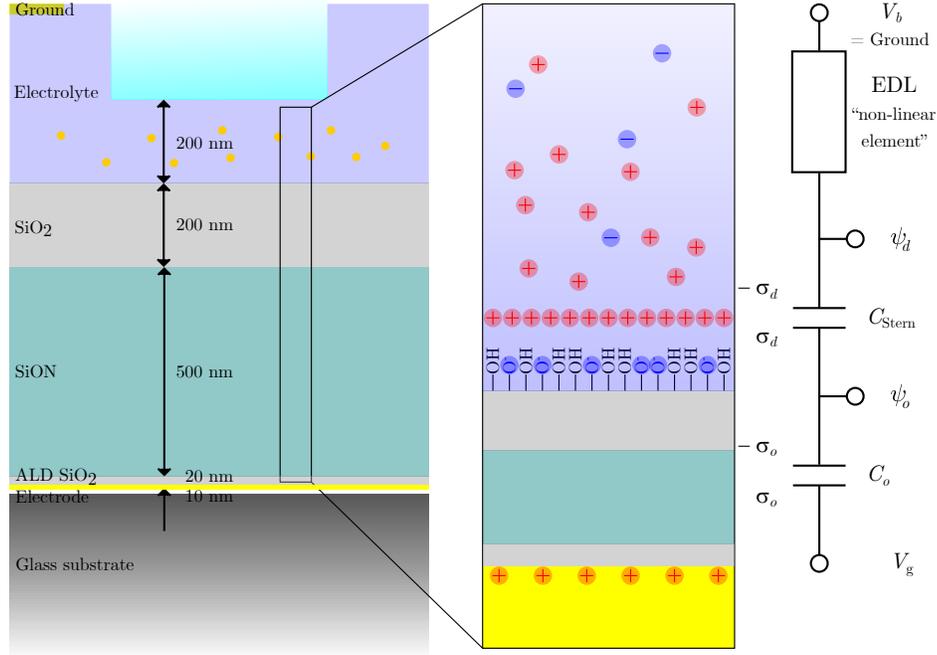

FIG. S9. Left: Sketch of the device cross section drawn to scale in the $z$-direction. Right: Blowup of the electrode-oxide-electrolyte region and the equivalent circuit. Same colors indicate the same layers in the left and middle panel. Not drawn to scale. Adapted from Jiang and Stein [S12].

The equivalent circuit model of Fig. S9 shows the voltage drop from the applied gate voltage at the electrode $V_\text{g}$ over a combination of two capacitors in series connected to a non-linear element. The first capacitor corresponds to the dielectric layer stack sandwiched between the electrode and the oxide surface. $SiO_2$ has a dielectric constant $\epsilon = 3.9$ [S13], while SiON can have dielectric constants between 3.9 and 7.5 [S13], depending on the relative proportion of N and O. The equivalent $SiO_2$ content can be estimated by measuring the optical properties of the material [S14]. In our case, ellipsometry measurements yielded a refractive index of $n_\text{SiON} = 1.64$, indicating a relative $SiO_2$ content of 64% and thus a dielectric constant of $\epsilon_\text{SiON} = 5.3$ [S14]. The capacitance per unit area of the dielectric stack is thus:

$$C_o = \left(\frac{1}{C_\text{SiON}} + \frac{1}{C_{SiO_2}}\right)^{-1} = \left(\frac{d_\text{SiON}}{\epsilon_0 \epsilon_\text{SiON}} + \frac{d_{SiO_2}}{\epsilon_0 \epsilon_{SiO_2}}\right)^{-1} \approx 5 \cdot 10^{-5}\,\text{F/m}^2. \tag{S12}$$

Therefore, the capacitively induced charge density on the oxide surface can be expressed as:

$$\sigma_o = C_o(V_g - \psi_o). \tag{S13}$$

The second capacitance $C_\text{Stern}$ is due to the presence of the Stern layer. It comprises a layer of ions adsorbed to the oxide surface. Because of the ions' finite site, the closest approach to the charged oxide surface is of the order of one ion radius. A capacitance arises between the charged oxide surface and the adsorbed ions. The measured Stern layer capacitance has been reported as [S15]

$$C_\text{Stern} = \frac{\sigma}{\psi_o - \psi_d} = 2.9\,\text{F/m}^2. \tag{S14}$$

Here, $\psi_d$ is the double layer potential.

### §2.5 Surface Chemistry

The behavior of the non-linear element of Fig. S9 is governed by the surface chemistry of the $SiO_2$ surface. The charge density $\sigma$ is related to the Stern layer capacitance by: [S12]

$$\sigma_d = C_\text{Stern}(\psi_o - \psi_d) = \frac{2\epsilon_0 \epsilon_r \kappa kT}{e} \sinh\left(\frac{e(\psi_d - V_b)}{2kT}\right), \tag{S15}$$

where the second equality is given by the Grahame equation. In our case, the bulk potential of the liquid $V_b$ is set to 0, as we ground the electrode in contact with the liquid.

The charge density $\sigma$ originates from the ionized silanol groups and follows [S8]

$$\sigma = \sigma_d - \sigma_o = -e\frac{\Gamma_\text{tot}}{1 + \frac{[H^+]}{10^{-pK}}}. \tag{S16}$$

Noting that $[H^+]_\infty = 10^{-pH}$ and making use of the Boltzmann relation to find the concentration of hydroxyl ions at the surface, $[H^+] = [H^+]_\infty \cdot \exp(-e(\psi_o - V_b)/kT)$, equation S16 can be rewritten as

$$\sigma = \frac{-e\Gamma_\text{tot}}{1 + 10^{(pK-pH)} \exp\left(-\frac{1}{kT}\left[\frac{\sigma + C_o V_g + C_\text{Stern}\psi_d}{C_\text{Stern} + C_o} - V_b\right]\right)}. \tag{S17}$$

Equation S17 is an implicit relation for the induced surface charge as a function of the system parameters. The value of $\psi_d$ is a second unknown and a second relation linking $\sigma$ to $\psi_d$ is therefore needed to solve the model. It is Gauss' equation in its interfacial form, for the present case:

$$\sigma = \frac{C_o}{(C_o^{-1} + C_\text{Stern}^{-1})^{-1}} \frac{2\epsilon_0 \epsilon \kappa kT}{e} \sinh\left(\frac{e(\psi_d - V_b)}{2kT}\right) + C_o(\psi_d - V_g). \tag{S18}$$

We now have two independent equations for two unknowns and can solve the model numerically: one first uses equation S18 to calculate $\sigma$ for a range of values of $\psi_d$; one then uses these value pairs to evaluate the right-hand side of equation S17. The unique solution for the two unknowns $\sigma$ and $\psi_d$ lies at the intersection between the two curves $\sigma(\psi_d)$ given by equations S17 and S18.

With the introduced model, one can now compute the response of a flat dielectric surface in contact with an aqueous electrolyte with a number density $\Gamma$ of chemical surface groups of dissociation constant $pK$, in contact with an electrolyte of a certain $pH$ and salt concentration characterized by a Debye length $\kappa^{-1}$.

### §2.6 Simulations of Electrostatic Fringe Fields

To extract the modulated surface potentials, we need to take into account the effect of the trap geometry on the electric field applied by the buried gate electrode. Indeed, for a sufficiently large trap, we do not expect the field across the dielectric to be disturbed. However, the smaller the trap, the more the electric field leaks into the trap, thereby reducing the expected modulation of surface potential. We performed finite-element simulations to quantify this fringing effect.

In the absence of space charge, the electrostatic Poisson equation simplifies to the Laplace equation

$$\tilde{\nabla}^2 \tilde{\psi} = 0, \tag{S19}$$

used here in its nondimensional form with $\tilde{\nabla} = \nabla/d$ and $\tilde{\psi} = \psi/V_g$. The permittivities throughout the dielectric stack are assumed equal. It is solved on a radially symmetric geometry in cylindrical $\tilde{r}$-$\tilde{z}$-coordinates, with the origin located on the gate electrode at the center of a circular hole with radius $r/d$, see Fig. S10. The Dirichlet boundary conditions (BCs) $\tilde{\psi}(\tilde{r} \geq r/d, \tilde{z} = 0) = 1$ and $\tilde{\psi}(\tilde{z} = 1) = 0$ are assigned to the gate electrode and the liquid respectively. At the hole in the gate electrode $\tilde{r} < 1$, $\tilde{z} = 0$, on the symmetry axis $\tilde{r} = 0$, and far from the hole $\tilde{r} \to \tilde{r}_\infty$, homogeneous Neumann BCs $\mathbf{n} \cdot \tilde{\nabla} \tilde{\psi} = 0$ were assigned, $\mathbf{n}$ being the normal vector. In order to eliminate finite-size effects, the domain size $\tilde{r}_\infty$ was chosen to be $10 r/d$ for $r/d > 1$ and 10 otherwise.

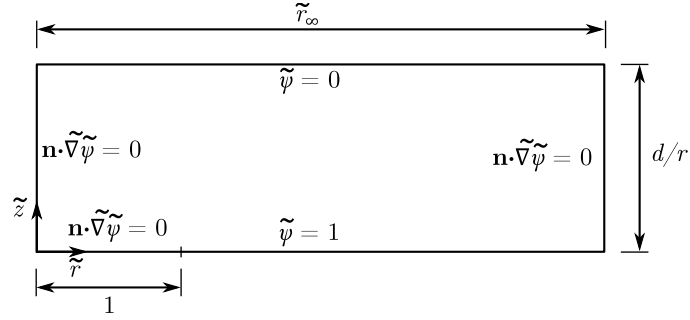

FIG. S10. Computational domain for the simulations of electrostatic fringe fields with dimensions and boundary conditions.

The problem was solved with the finite-element code COMSOL Multiphysics version 6.0, using quadratic shape functions, the direct MUMPS solver, and a structured rectangular grid. To ensure grid independence of the results, the grid was systematically refined twice and the normal electric field directly above the hole, $-\partial_z \tilde{\psi}(\tilde{r} = 0, \tilde{z} = 1)$, was used as a representative value for grid assessment. A Richardson extrapolation was performed to find the grid-independent value and the final grid ensures a relative deviation from this value of $\leq 0.002$. Note that the simulated geometry used planar boundaries at $\tilde{z} = 0$ and $\tilde{z} = 1$. Neglecting the recesses of a depth $\approx 10\,\text{nm}$ on the $\approx 700\,\text{nm}$ thick glass layer introduces a $< 2\%$ error in the geometry and thus in the calculated electric fields.

Finally, the non-dimensional normal electric field at the glass-water interface was evaluated for different hole sizes $r/d$:

$$\mathrm{E}_g(\tilde{r})/(V_g/d) = -\partial_z \tilde{\psi}(\tilde{r}, \tilde{z} = 1). \tag{S20}$$

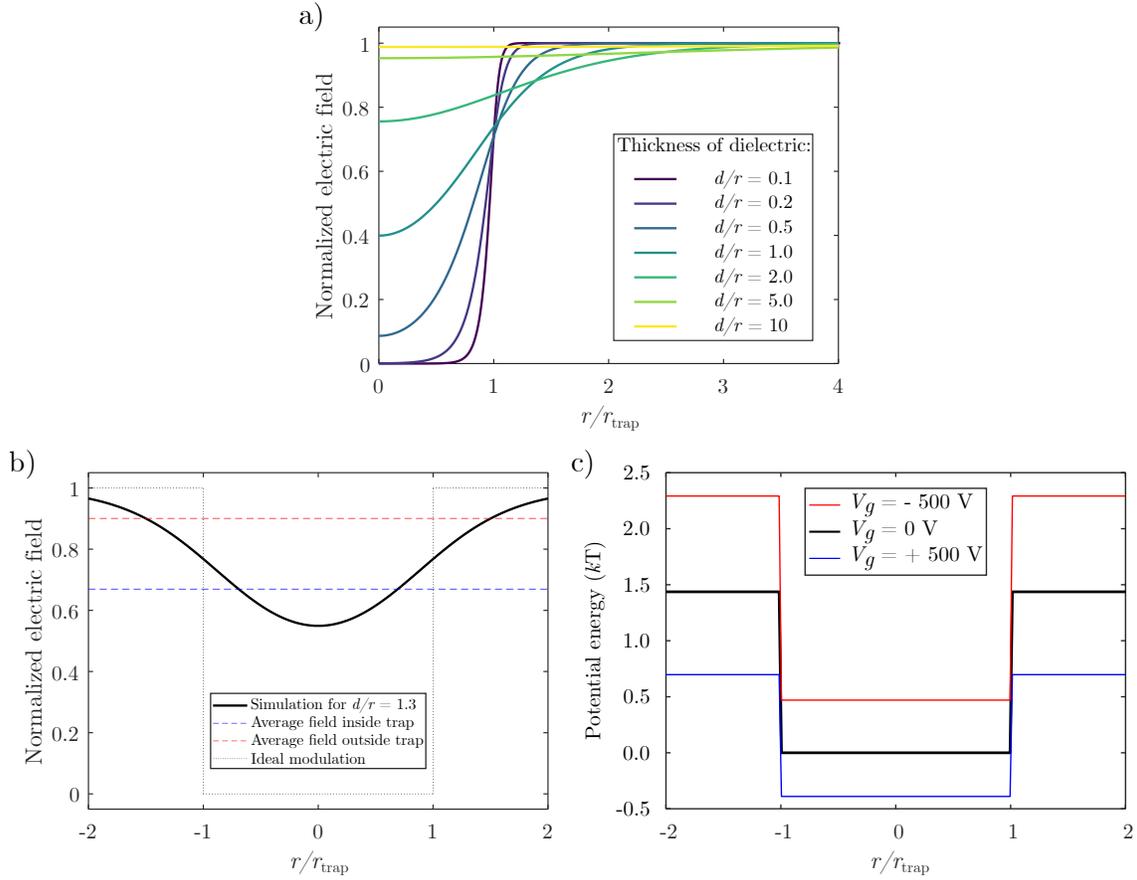

FIG. S11. a) Numerical simulations of the effect of the dielectric thickness on the normal component of the electric field developing in the hole. As the dielectric gets thicker, the modulation is less and less pronounced at the hole location. b) In our experiments, the $d/r$ ratio is 1.3. The average field inside the trap is depicted as blue dotted line and is 0.67 in units of the electric field across the dielectric stack far from the traps. As the traps are arranged at distances of $4r_{\text{trap}}$, the field outside the traps does not reach 1. Averaged over the region outside the traps, the mean normalized electric field is 0.9. c) The consequence on the potential well depth is schematically depicted: the surface potential in the trap area will also be modulated, though less than the region outside the trap. Levels show the averaged potential energy within and outside of the traps to visualize the potential well as a representative rectangular well. From the reference situation depicted as black solid line, where we set the trap bottom to the reference potential energy, both the trap bottom and the area surrounding the trap increase (decrease) upon applying a negative (positive) gate bias.

* fparatore@ethz.ch  
† ark@zurich.ibm.com

[S1] F. Paratore, V. Bacheva, G. V. Kaigala, and M. Bercovici, Proceedings of the National Academy of Sciences **116**, 10258 (2019).
[S2] S. Fringes, *Transport and Assembly of Nanoparticles in a Tunable Nanofluidic Confinement* (PhD Thesis, Department of Chemistry, University of Zurich, 2017).
[S3] F. Ruggeri, A. Ozyhar, M. Rozycka, M. Wojtas, F. Zosel, B. Schuler, M. Krishnan, and N. Mutter, Nature Nanotechnology **12**, 488 (2017).
[S4] A. Apelblat and J. Barthel, Zeitschrift für Naturforschung A **46**, 131 (1991).
[S5] G. H. Zimmerman, H. Arcis, and P. R. Tremaine, Journal of Chemical and Engineering Data **57**, 2415 (2012).
[S6] S. Fringes, F. Holzner, and A. W. Knoll, Beilstein Journal of Nanotechnology **9**, 301 (2018).
[S7] G. M. Bell, S. Levine, and L. N. Mccartney, Journal of Colloid and Interface Science **3**, 335 (1970).
[S8] S. H. Behrens and D. G. Grier, The Journal of Chemical Physics **115**, 6716 (2001).
[S9] B. Derjaguin, Kolloid-Zeitschrift **69**, 155 (1934).
[S10] P. Nicollier, C. Schwemmer, F. Ruggeri, D. Widmer, X. Ma, and A. W. Knoll, Physical Review Applied **15**, 034006



\end{document}